\documentclass[twocolumn]{svjour3}

\RequirePackage[T1]{fontenc}

\smartqed  

\RequirePackage{graphicx}
\RequirePackage{mathptmx}      
\RequirePackage{flushend}
\RequirePackage[numbers,sort&compress]{natbib}
\RequirePackage[colorlinks,citecolor=blue,urlcolor=blue,linkcolor=blue]{hyperref}

\RequirePackage[utf8]{inputenc}
\RequirePackage{amssymb}
\RequirePackage{amsmath}
\RequirePackage{braket}
\RequirePackage{xcolor}

\journalname{Eur. Phys. J. B}

\begin{document}

\title{Time delay in 1D disordered media with high transmission}

\author{Luis A. Razo-L\'opez
\and
J. A. M\'endez-Berm\'udez
\and
Victor A. Gopar
}

\institute{
Luis A. Razo-L\'opez \at Université Côte d'Azur, CNRS, Institut de Physique de Nice (INPHYNI), 06100 Nice, France
\and
J. A. M\'endez-Berm\'udez \at Instituto de F\'{\i}sica, Benem\'erita Universidad Aut\'onoma de Puebla, Apartado Postal J-48,  72570, M\'exico
\and
Victor A. Gopar \at Departamento de F\'isica Te\'orica, Facultad de Ciencias, and BIFI,  Universidad de Zaragoza, Pedro Cerbuna 12, ES-50009 Zaragoza, Spain
}

\date{\today}

\maketitle

\begin{abstract}
We study the time delay of reflected and transmitted  waves in 1D disordered media with high transmission. 
Highly transparent and translucent  random media are found in nature or can be synthetically produced. We perform numerical simulations of microwaves propagating in disordered waveguides to show  that reflection amplitudes are described by complex Gaussian random variables with the remarkable consequence that the time-delay statistics in reflection of 1D disordered media are described as in random media in the diffusive regime. For transmitted waves,  
we show numerically that the time delay is an additive quantity and its fluctuations thus follow a Gaussian distribution. Ultimately,  the distributions of the time delay in reflection and transmission are physical illustrations of the central limit theorem at work. 
\end{abstract}

\section{Introduction}

Classical and quantum waves in disordered media such as microwaves propagating in a random waveguide and electrons passing through quantum wires, respectively, are delayed relative to waves traveling in free space. 
Apart from the fundamental question about how much time waves spend in a random medium, the delay provides practical information. On a macroscopic scale, for instance, the time delay is an essential ingredient in 
the construction of medical images from the reflected ultrasound waves penetrating the body; while on a microscopic scale, in electronic quantum transport, the time delay is directly related with the density 
of states and the  admittance of  quantum capacitors~\cite{Buttiker_1993,Buttiker_1996,Christen_1996, Brouwer_1997, Gopar}. See also \cite{Carvalho}. A related useful quantity to the time delay in transmission and reflection is the dwell time, which has been used to find the position of a reflector embedded in a disordered structure~\cite{Yiming_2021}. 
Thus,  the time delay has been an issue of interest in  fundamental and practical investigations in  different research areas. 

In real materials, the disorder is ubiquitous and is a source of
random fluctuations of static and dynamical properties, such as the transmission and the time delay. 
For coherent wave propagation,  localization effects on the random fluctuations of time delay have been of particular interest. 
Despite many efforts to describe theoretically the time delay in disordered media, a complete theoretical description of this problem is still missing, even for the simplest case of 1D systems. However, significant advances have been made when imposing some restrictions. For instance,  
several works consider the time delay of disordered systems with an infinite potential at the end of the samples; thus, the statistics of the time delay of reflected waves has been  obtained in the localized and ballistic regimes \cite{Texier_1997,Texier_1999}. Further   progresses have been made in the description of the time delay in reflection and transmission by considering large disorder samples in the insulating transport regime \cite{Bolton_1999, Schomerus2001}. Under these restrictions an extensive literature exists.  For a review of the topic, which includes the case of disordered systems beyond 1D, see for instance Refs.~\cite{Texier2016,Kottos2005}. 

In this paper, we study the time delay of reflected and transmitted waves in 1D disordered structures composed of weak scatterers such that the mean free path $\ell$ is larger than the sample length $L$. The total transmission coefficient through the samples is thus near-unity.
In the opposite insulating regime ($L \gg \ell$), the time delay in reflection and transmission in 1D has been studied in Refs. \cite{Texier_1999} and \cite{Schomerus2001}, respectively. 

Because of the large random fluctuations of the time delay, we are particularly interested in the complete distribution of those fluctuations. As previously pointed out, the time delay is a quantity of interest in classical and quantum waves. Here we will use numerical simulations of microwaves in a waveguide with randomly placed scatterers to show that the random reflection amplitudes are complex Gaussian random variables and as a consequence, the distribution of the time delay of reflected waves follows a probability density function,  Eq.~(\ref{poftaur}), theoretically and experimentally obtained in  Ref.~\cite{Genack1999}, albeit considering systems in the diffusive regime. For the time delay in transmission, we show numerically that it is an additive quantity and thus the distribution of the time delay follows a Gaussian distribution according to the central limit theorem (CLT).

Transparent and translucent disordered media characterized by high transmission coefficients 
can be found in nature, such as organic materials where light can easily pass through. Also,
synthetical materials like  polycrystalline ceramics and optical fibers with high transmission are manufactured nowadays~\cite{Kim,Mafi_2019}. Furthermore, 
though the presence of the disorder is generally considered as a disadvantage, recent 
experimental technics have exploited the presence of disorder  
to increase the transparency  of polycrystalline ceramics, to improve focusing in semi-transparent media, and enhancing energy transport~\cite{Hsieh,Kim_2012,Battista,Pattelli,Osipov,Kholoud,Mafi_2019,Vellekoop,Shi}. It has also been shown numerically and experimentally that disorder in periodic multilayer structures can be used as a broadband source and increase the x-ray reflectivity \cite{Loevezijn}. 

\begin{figure}
\includegraphics[width=\columnwidth]{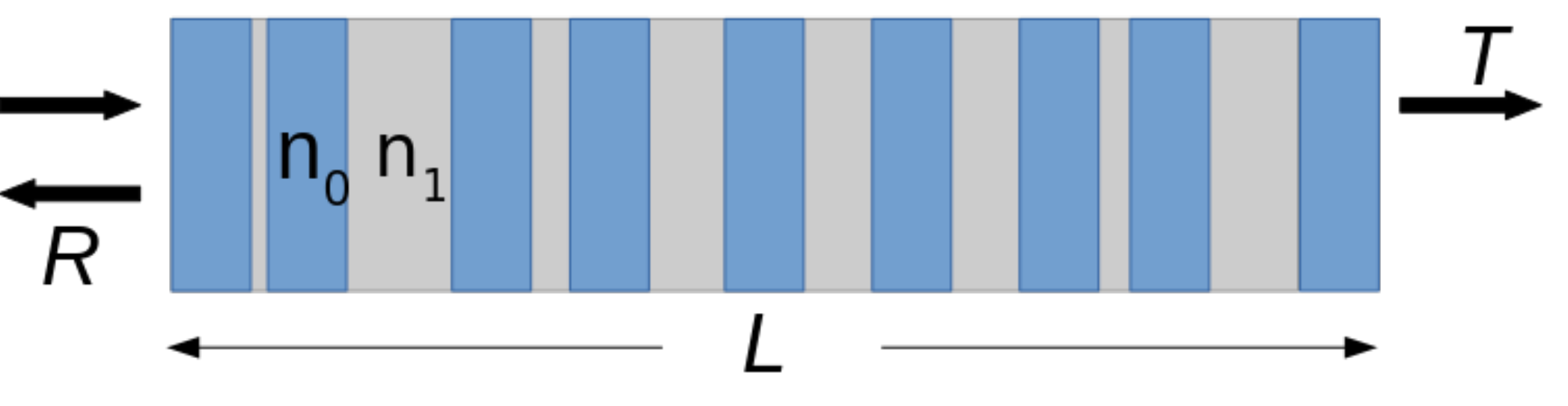}
\caption{A schematic of the waveguide consisting of alternating layers with indices of refraction $n_0$ and $n_1$.}
\label{fig_1_resub}
\end{figure}

Let us start considering a 1D disordered sample, as shown in Fig. \ref{fig_1_resub}, and assume that the scattering matrix 
$S$ that relates incoming and outgoing waves at  both sides of the system is given by  
\begin{eqnarray}
\label{Smatrix}
{S} & = & \left(\begin{array}{cc}
r & t  \\
t  & r' \\
\end{array} \right), 
\end{eqnarray}
where the complex numbers $r$ and $r'$ are the reflection amplitudes at the right and left sides of the system, respectively, and $t$ is the transmission amplitude. The elements of the $S$-matrix are frequency  dependent.  In writing Eq.~(\ref{Smatrix}), we have assumed time-reversal invariance, so the  transmission of incident waves from the left and right sides of the sample are the same. Following seminal  works by  Wigner and Smith \cite{Wigner1955,Smith1960}, the time delays  of reflected  and transmitted  waves are determined  by  the derivative of the argument of the reflection and  transmission  amplitudes, $\theta_r$ and $\theta_t$, with respect to the frequency $\omega$: $\tau_r = d\theta_r/d\omega$ and
$\tau_t = d\theta_t/d\omega$.

We obtain the scattering matrix $S$ numerically using the transfer matrix method  described in Ref.~\cite{Markos}. The method consist of obtaining the total transfer matrix $M$ of the whole sample which is given by the product of individual transfer matrices $M_i$ associated to each  layer. See also Ref.~\cite{Razo2020}:
\begin{equation}
 M =M_1 M_2\cdots M_N=  \left(\begin{array}{cc}
1/t^* & -r^*/t^*  \\
-r/t  & 1/t \\
\end{array} \right), \nonumber
\end{equation}
where $N$ is the number of layers. 
Thus, the elements of the scattering matrix $S$ are obtained from the transfer matrix $M$. 
Microwaves at the characteristic frequency $\nu=8$~GHz ($\omega=50$~rad$\cdot$ns$^{-1}$) are launched into a multilayer structure  with alternating indices of refraction $n_0=1$ (air) and $n_1=1.05$, see Fig. \ref{fig_1_resub}. The widths of the 
air-layers are normally distributed $\mathcal{N}(0,1)$, while the $n_1$-layers are of width $0.01$~cm. We generate an ensemble of $10^6$ samples with different realizations of the disorder and collect the data of the reflection and transmission amplitudes across the ensemble. That allows us to obtain different statistical properties such as the ensemble average of the logarithmic transmission $\langle \ln T  \rangle$, from which the mean free path 
$\ell$ is obtained via the relation $\langle -\ln T \rangle = L/\ell$ \cite{Mello_book}. With the above parameters, we find that $\ell \approx 2.5\times10^{6}$~cm. Therefore, samples are  weakly disordered ($k\ell\gg1$). We remark that coherent multiple scattering by weak scatterers produce random fluctuations of the phases $\theta_r$ and $\theta_t$ \cite{Stone_1983}. Thus $\tau_r$  and $\tau_t$ 
are also random variables and a statistical analysis of their   fluctuations is necessary. The 
time delays are obtained using a differential frequency step $\Delta \nu=47.71$~Hz ($\Delta \omega= d\omega=3 \times10^{-7}$~rad$\cdot$ns$^{-1}$). 
In the following sections, we will obtain the distribution of the fluctuations of both $\tau_r$ and $\tau_t$.

\section{Time delay of reflected waves}

A key point to obtain the complete distribution of $\tau_r$ is to notice that 
the reflection amplitude $r(\omega)$ is a complex Gaussian 
random variable. 
Furthermore, at two frequencies, say $\omega_1$ and $\omega_2$, the joint  distribution of $r_1=r(\omega_1)$ and $r_2=r(\omega_2)$,  
also follows a Gaussian distribution. 
We point out that the limit Gaussian distribution of $\bf r$ is a consequence of 
the CLT when the random vector ${\bf r}$ is interpreted as the result of the sum of a large number of complex random variables or complex phasors ~\cite{Goodman}. 

Defining  $r$ as the column matrix ${\bf r}=(r_1,r_2)^T$ and assuming circular symmetry, the joint distribution $P({\bf r, r^{\dagger}})$ can be written as~\cite{Goodman}
\begin{equation}
 \label{Gaussian}
 P({\bf r,r^\dagger})=\frac{1}{ \pi^2 \det C} \exp{\left(-{\bf r}^\dagger C^{-1} {\bf r} \right)}, 
\end{equation}
where $C_{ij}=E[r_ir_j^*]$ is the Hermitian covariance matrix and ${\bf r}^\dagger$ is the 
conjugate transpose of $\bf r$.  Circular symmetry  implies the following relations between the expectation values: $E[\mathrm{Re}(r_1)]=E[\mathrm{Im}(r_1)]=0$, $E[\mathrm{Re}(r_1)\mathrm{Re}(r_2)]=E[\mathrm{Im}(r_1)\mathrm{Im}(r_2)]$, and $E[\mathrm{Re}(r_1)\mathrm{Im}(r_2)]=-E[\mathrm{Im}(r_1)\mathrm{Re}(r_2)]$.

\begin{figure}
\includegraphics[width=\columnwidth]{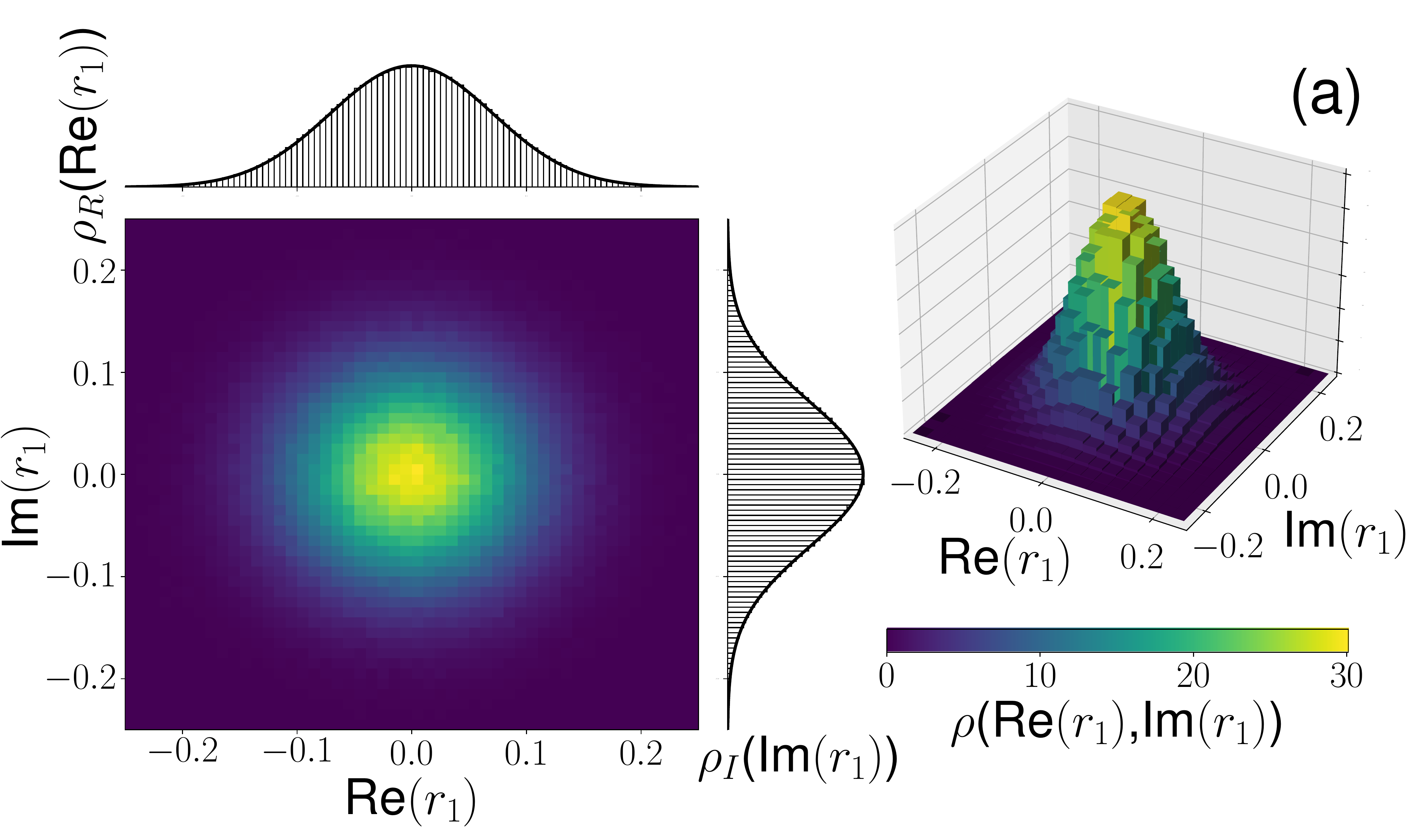}
\includegraphics[width=\columnwidth]{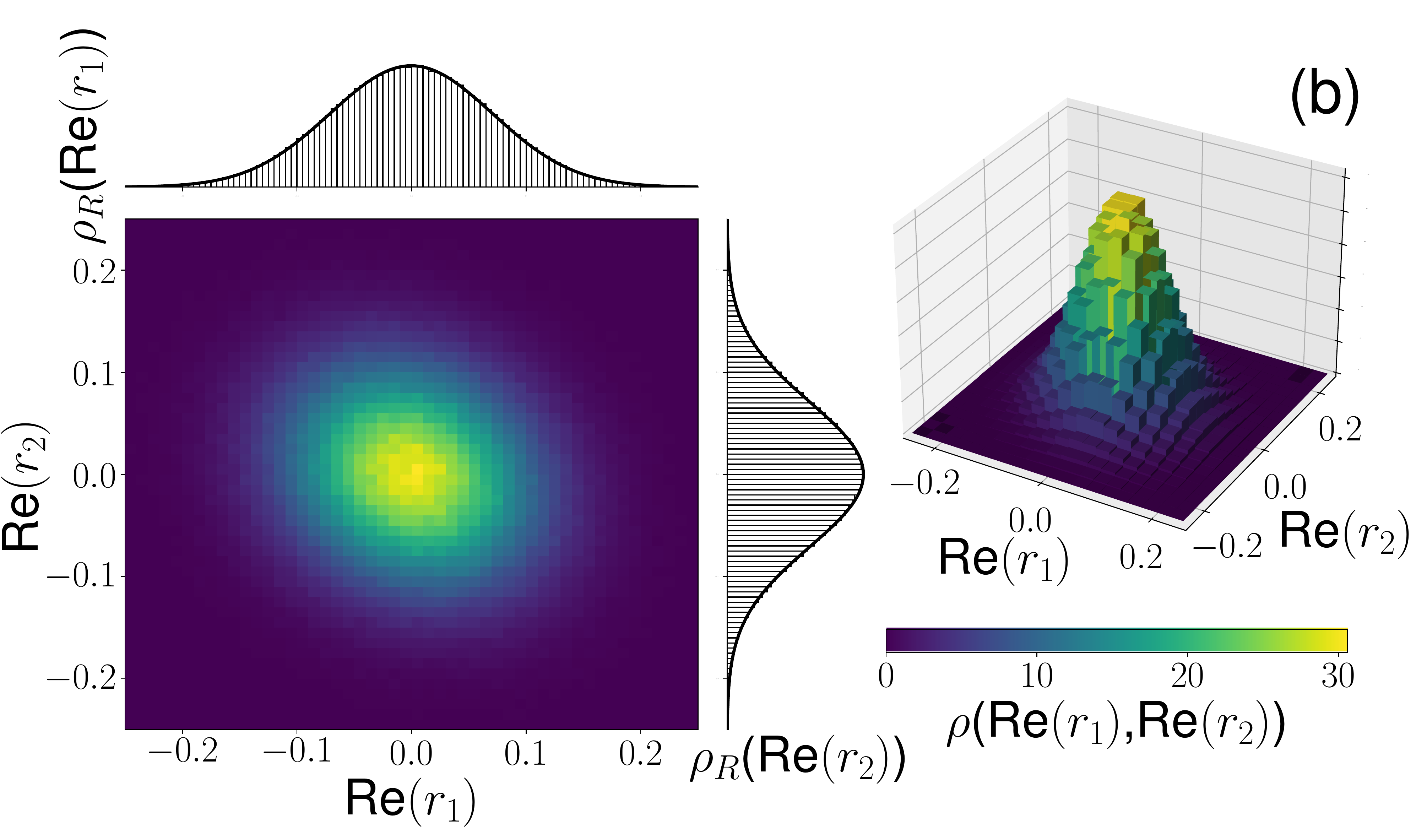}
\caption{
 Numerical joint density distribution of (a) the real and imaginary parts of the reflection amplitude $r$ and (b) the real parts of the reflection amplitude $r$ at two frequencies: $\omega_1=50.262$~rad ns$^{-1}$ and $\omega_2=50.268$~rad ns$^{-1}$.  Panels at the left present a top view of the joint distributions and their respective marginal Gaussian distributions in gray. In both examples (a) and (b), $L/\ell=10^{-2}$.}
\label{fig_2_resub}
\end{figure}

We illustrate the features of the circular Gaussian  variables using the  numerical simulations of disordered structures, described above. In Fig.~\ref{fig_2_resub}(a) (right), we show the joint density distribution of the real and imaginary parts of the reflection amplitudes, 
$\rho(\mathrm{Re}(r_1),\mathrm{Im}(r_1))$, and a top view (left) of this density. 
Additionally, in  Fig.~\ref{fig_2_resub}(a) (left), the marginal Gaussian density distributions of the real and imaginary parts  $\rho_R(\mathrm{Re}(r_1))$ and $\rho_I(\mathrm{Im}(r_1))$, respectively, are plotted in gray.

The joint density distribution of the real part of the reflection amplitudes at two frequencies $\rho\left(\mathrm{Re}(r_1),\mathrm{Re}(r_2)\right)$ is shown in Fig.~\ref{fig_2_resub}(b) (right), while the left side panel shows a top view of this density.  
In the same Fig.~\ref{fig_2_resub}(b) (left), we plot in gray the  marginal Gaussian  distributions $\rho_R(\mathrm{Re}(r_1))$ and $\rho_R(\mathrm{Re}(r_2))$. Thus, the reflection amplitudes of our disordered structures are circular random Gaussian variables. 
\begin{figure}
\includegraphics[width=\columnwidth]{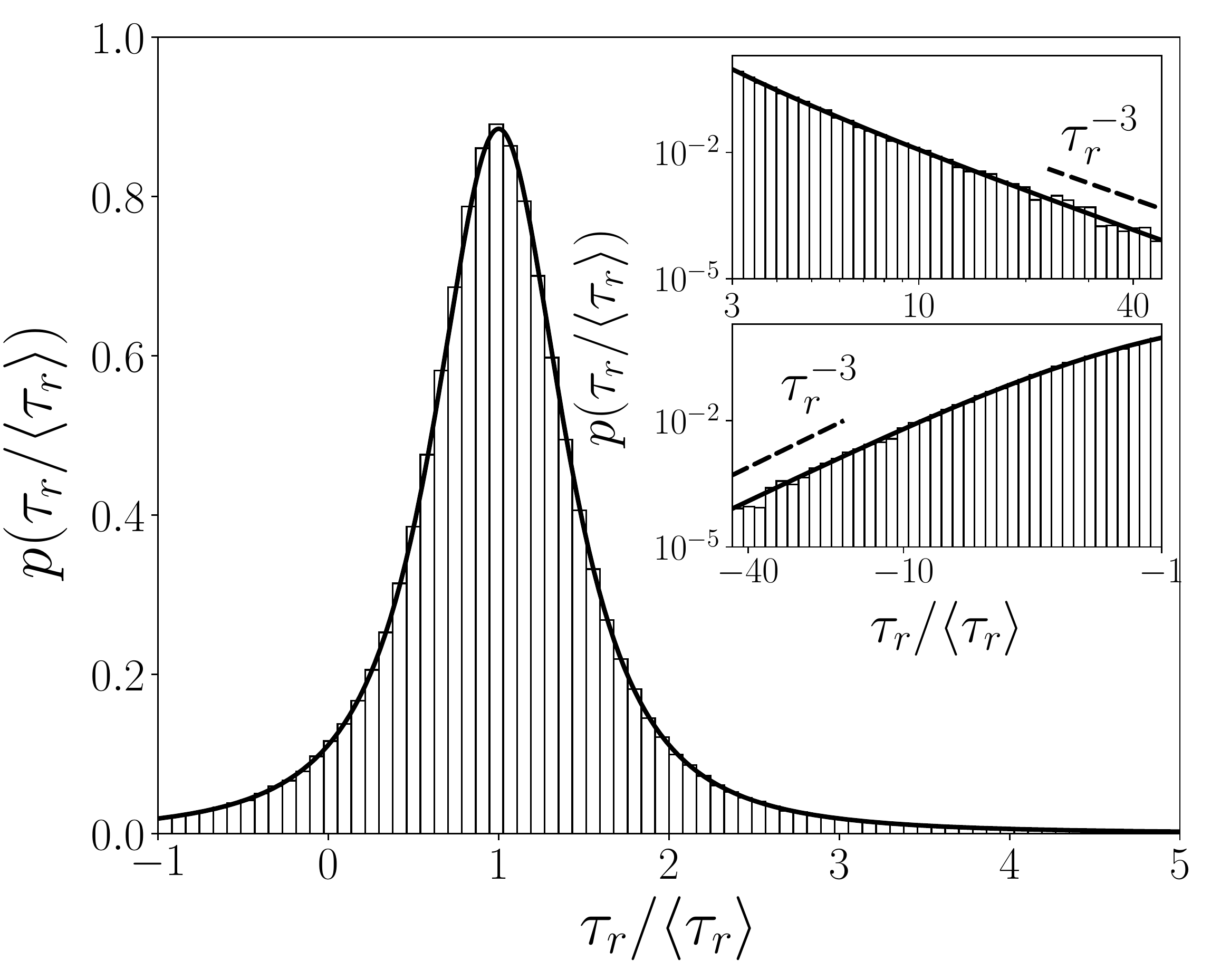}
\caption{
The numerical distribution $p(\tau_r/\langle \tau_r \rangle)$ (histogram) with $L/\ell=10^{-2}$ and $\langle \tau_r \rangle=4.35 \times 10^{-5}$ s. Solid curves are the theoretical prediction from Eq.~(\ref{poftaur}) normalized by the mean value $\tau_r\rightarrow\tau_r/\langle \tau_r \rangle$. The value $Q=0.335$ was obtained from the numerical simulation. The inset shows $p(\tau_r/\langle \tau_r \rangle)$ in a logarithmic scale. The dashed line, following the power law $\tau_r^{-3}$, is a guide to the eye.}
\label{fig_3_resub}
\end{figure}

Statistical properties of 
waves in random media have been studied assuming circular random Gaussian variables, as described by Eq. (\ref{Gaussian}) 
\cite{Rice,Genack1999}. See also Ref.~\cite{Schomerus2001}. In particular, the distribution of the frequency derivative of the phase of the reflection has been theoretically and experimentally obtained in Refs.~\cite{Genack1999, Tiggelen1999} in the  diffusive regime defined by: $\ell \ll L \ll \xi$, where $\xi$ is the localization length. For the time delay of localized waves 
see Ref.~\cite{Chabanov2001}.  Here,  we are dealing with 
structures in a different regime, i.e.,  $L < \ell (=\xi/2, \mathrm{\ \ for\ \ 1D \ \ systems})$; however, as we have shown above, the reflection is well described by a circular complex Gaussian random process. Therefore the distribution of the time delay $p(\tau_r)$ can be written as \cite{Genack1999,Tiggelen1999}:
\begin{equation}
 p(\tau_r)=\frac{1}{2\langle \tau_r \rangle}\frac{Q}{\left[Q+(\tau_r/\langle \tau_r \rangle-1)^2\right]^{3/2}},
 \label{poftaur}
\end{equation}
where $Q=-2b/a^2-1$ with $a=\langle \tau_r \rangle$ and $b$ is  proportional to $\langle \tau_r \rangle^2$. To compare the distribution in Eq. (\ref{poftaur}) with numerical simulations,   we obtain the  
constants $a$ and $b$ from the covariance element $C_{12}$ 
as $a=\mathrm{Im}\left({C_{12}}\right)/\Delta \omega$ and $b=\mathrm{Re}\left({C_{12}}-1\right)/(\Delta \omega)^2$. The elements $C_{ij}$ have been  normalized, $C_{ii}=1$. 

We show in Fig.~\ref{fig_3_resub} the distribution of the time delay in reflection $p(\tau_r)$ as given by Eq.~(\ref{poftaur}) (continuous line) and the corresponding  numerical distribution (histogram) for disordered structures with $L/\ell=0.01$. A good agreement is seen. We remark that there are no fitting parameters. All the information about the parameters that enter into Eq. (\ref{poftaur}), $a$ and $b$, is extracted from the numerical simulations.

In the insets of Fig.~\ref{fig_3_resub}, the left and right tail of  $p(\tau_r)$ are shown on a log scale for a better appreciation of  
the power-law decay $\tau_r^{-3}$ of the distribution, top and bottom insets, respectively. This power-law behavior has also been observed experimentally and numerically in the diffusive regime~\cite{Genack1999,Tiggelen1999}.

\begin{figure}
\includegraphics[width=\columnwidth]{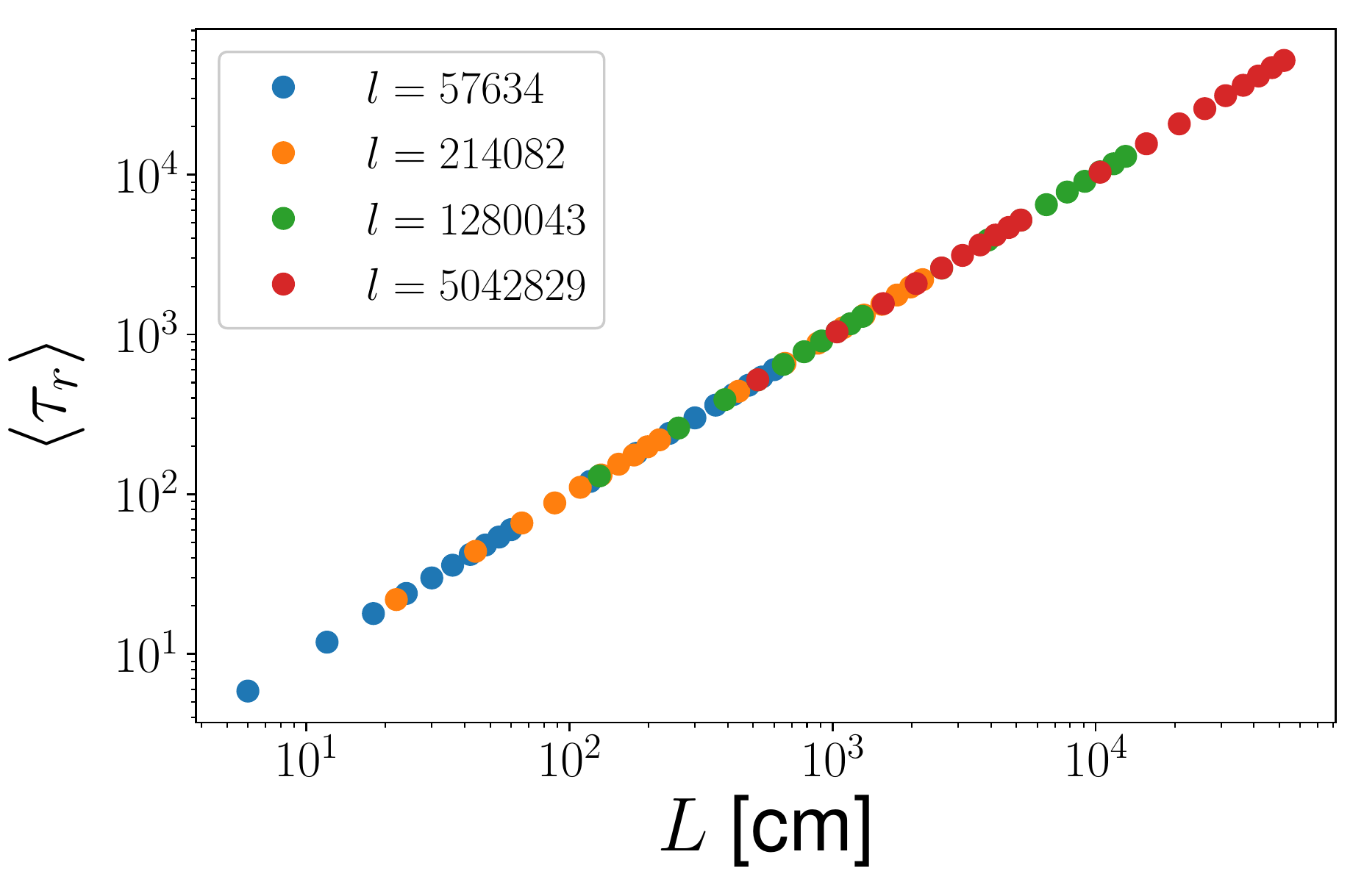}
\caption{
The average of the time delay in reflection $\langle \tau_r \rangle$ as a function of the system length $L$ for differents strength of the disorder characterized by the mean free path $\ell$ (reported in cm).}
\label{fig_4_resub}
\end{figure}

To gain some insight into the evolution of the statistics of the time delay with the system length, we study the first and second moments of $\tau_r$ as a function of the system length.
The average $\langle \tau_r \rangle$ is given by~\cite{Texier2016,Razo2020,Yiming_arxiv}
\begin{equation}
\label{avertau}
 \langle \tau_r \rangle = L n_e/c,
\end{equation}
where $n_e$ is the effective index of refraction and $c$ the speed of light. 
Notice that the average time delay depends only on the system's length, for $n_e$ fixed. This result agrees with a general property: the average time delay is determined essentially by the boundary of the system and it is independent of the details of the disorder. This interesting property has been studied experimentally and theoretically  ~\cite{Pierrat2014,Savo2017,Razo2020,Yiming_arxiv, Tiggelen2017}, and it has been deduced based on the direct relation of the average of the total time delay, $\langle \tau \rangle=\langle T \tau_t \rangle +\langle R \tau_r \rangle$, with the density of states inside the sample~\cite{Friedel,Krein, Tiggelen2017}. Notice that in our 1D systems, as a consequence of the unitarity of the scattering matrix,  $\langle \tau_t \rangle=\langle \tau_r \rangle$ \cite{Schomerus2001}. To illustrate the independence of $\langle \tau_r \rangle$ on the details of the disorder, in Fig. \ref{fig_4_resub} we plot the average $\langle \tau_r \rangle$ as a function of the length of the system for different values of the mean free path but keeping approximately constant the value of $n_e$. The mean free path can be considered a measured of the strength of disorder. As it can be seen, the value  $\langle \tau_r \rangle$ does not change with the mean free path.

\begin{figure}
\includegraphics[width=\columnwidth]{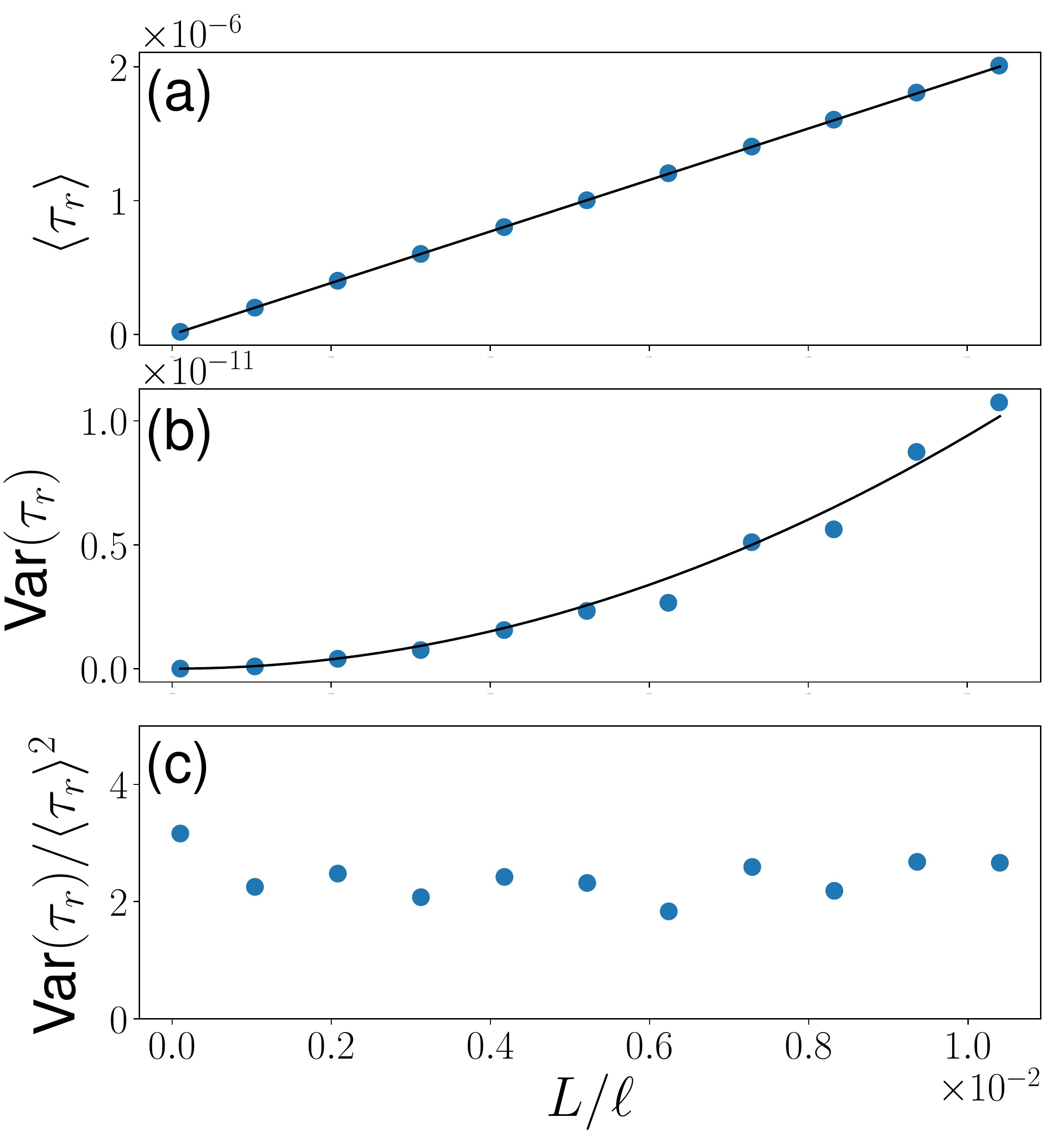}
\caption{(a) Numerical average of the reflection delay-time $\langle \tau_r \rangle$ for disordered structures with  $\ell \approx 57635$~cm (dots). The black continuous line is given by Eq.~(\ref{avertau}). (b) Numerical variance of the reflection delay-time Var$(\tau_r)$ (dots). The black continuous line is a fitting of $f(L)=aL^2$ with $a\approx 2.8\times 10^{-17}$. (c) The ratio 
$\mathrm{Var}(\tau_t)/\langle \tau_t \rangle^2$ as a function of the system length. The ratio remains approximately constant showing that  $\tau_r$ does not self-average.}
\label{fig_5_resub}
\end{figure}

Concerning the variance Var$(\tau_r)=\langle \tau_r^2 \rangle- \langle \tau_r \rangle^2$, we have obtained  a quadratic behavior with $L/\ell$. See Fig.~\ref{fig_5_resub}(b). In Fig.~\ref{fig_5_resub}, we kept $\ell$ fixed. In Fig.~\ref{fig_5_resub}(a), it is shown the linear behavior of $\langle \tau_r \rangle$ with $L$, as discussed above. We notice, however, that the power-law tail of $p(\tau_r)$ in Eq.~(\ref{poftaur}) leads 
to the divergence of the second moment $\langle \tau_r^2 \rangle$ 
when considering the whole number line where $p(\tau_r)$ is supported. This divergence is suppressed by the finite frequency step $\Delta w$ in our numerical simulations.  But still, large fluctuations of $\tau_r$ are expected.  
For instance, we notice a non smooth behavior of the variance  in Fig.~\ref{fig_5_resub}(b). The strong fluctuations of $\tau_r$ are better revealed by plotting the ratio $\mathrm{Var}(\tau_r)/ \langle \tau_r \rangle^2$, as in Fig. \ref{fig_5_resub}(c). 
As one can see, this ratio does not decrease with the length, but instead it  
 remains approximately constant with the system length.  This means that $ \tau_r $ is non self-averaging quantity. In the next section, we will contrast this result with the time delay in transmission. 
 
\section{Time delay of transmitted waves}

We now consider the time delay $\tau_t$ of waves that pass through the whole sample. We recall that there is a high probability that waves can travel through the entire samples  without being reflected since the transmission coefficient of the samples is close to unity. 
In this case, we found that the time delay in transmission is an additive quantity.  That is, 
the delay time in transmission  can be seen as the sum of partial time delays produced  by each scatterer or layer of the sample. 

\begin{figure}
\includegraphics[width=\columnwidth]{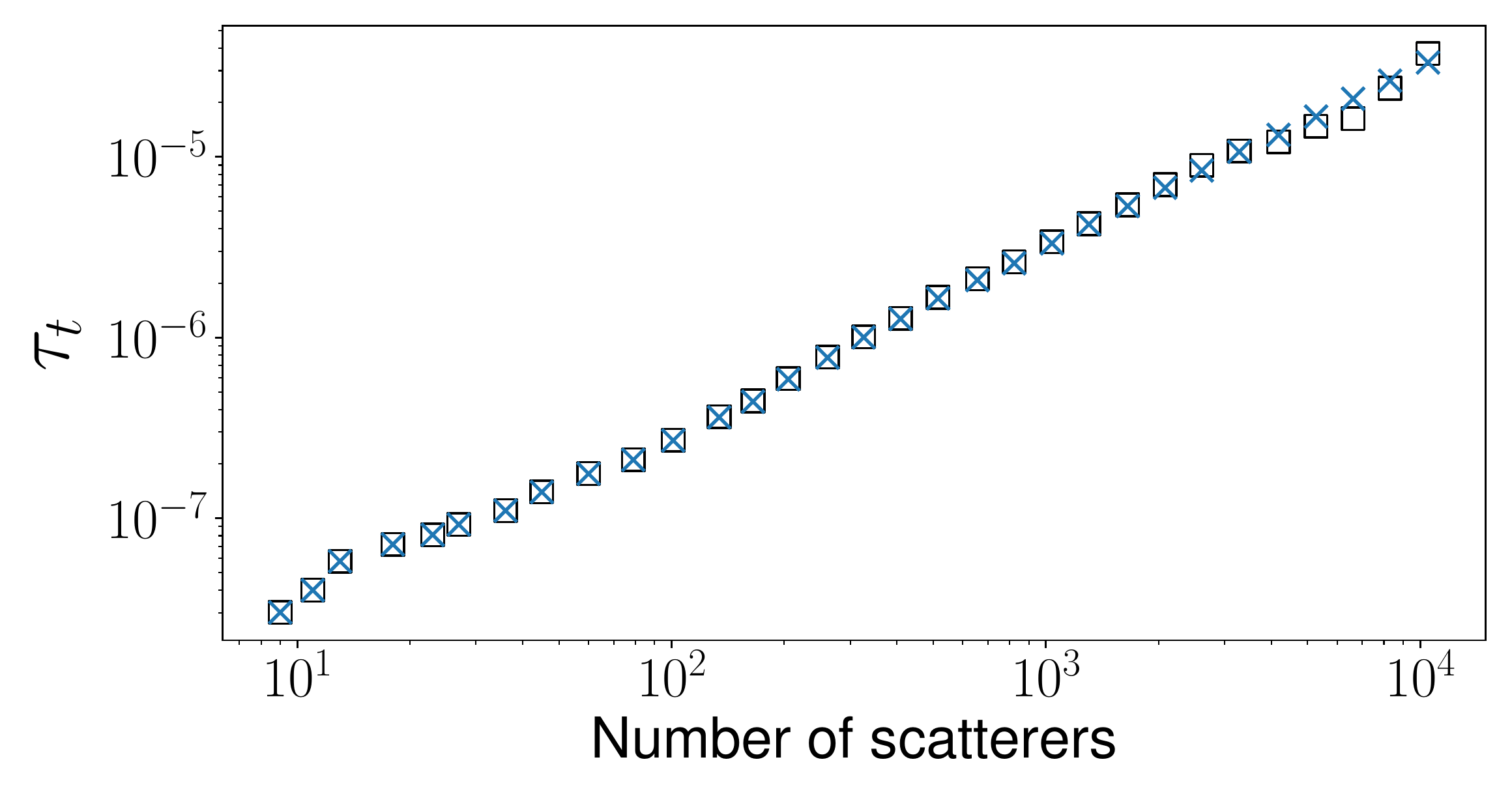}
\caption{The time delay of transmitted waves $\tau_t$ as a function of the number of scatterers for a typical sample of $\ell=9480$ cm.  Crosses depict the results obtained by adding partial contributions of each scatterer to the time delay,  while squares correspond to the time delay obtained from the derivative of the total cumulative phase of the transmitted wave.}
\label{fig_6_resub}
\end{figure}

We illustrate the additive property of the time delay in transmission $\tau_t$ using typical disordered samples. In  Fig.~\ref{fig_6_resub}, we show the results of $\tau_t$ as a function of the number of scatterers in the sample. 
Squares  are obtained from the frequency derivate of the phase of the amplitude transmission $t$ through the entire sample. In contrast, crosses show the values of $\tau_t$  obtained from the sum of time delays associated with  each scatterer, i.e.,  $\tau_t=\sum_{i=1}^N {\tau_t}_i$.

As we can see in Fig.~\ref{fig_6_resub}, the results of both procedures to obtain $\tau_t$ agree. Some  differences can be observed for a large number of scatterers, when the transmission coefficient starts to deviate from unity.

Consequently, the average time delay is proportional to the number of slabs or, equivalently, to the length of the samples. Now, since 
the time delay in transmission is given by the sum of independent random variables, we expect from the CLT that the probability density of $\tau_t$ follows a Gaussian distribution, as we verify next.
\begin{figure}
\includegraphics[width=\columnwidth]{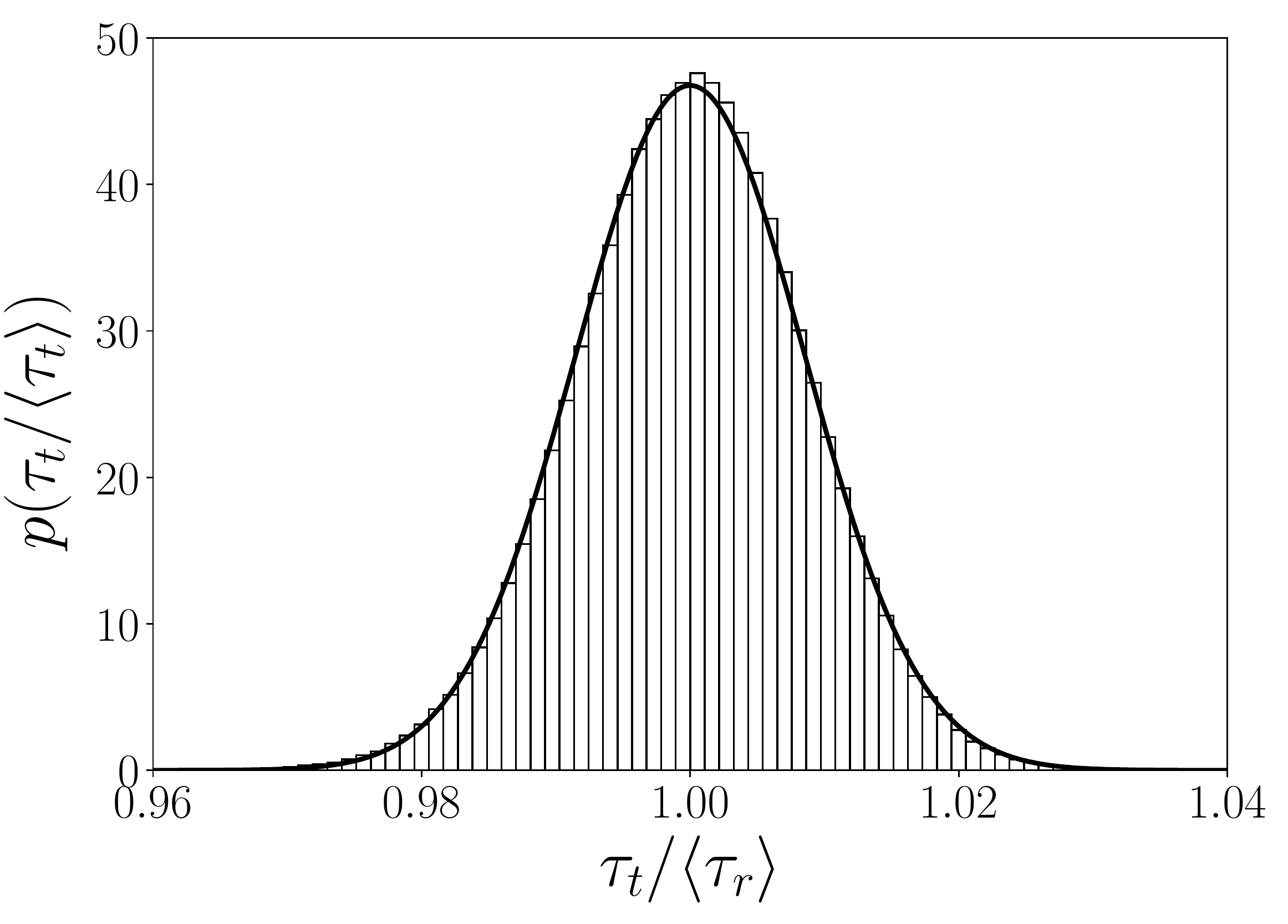}
\caption{The numerical distribution $p(\tau_t)$ (histogram) for disordered structures characterized by  $L/\ell\approx 10^{-2}$ and $\langle \tau_r \rangle=4.35\times 10^{-5}$ s. The continuous curve is a Gaussian where $\mu$ and $\sigma^2$ have been extracted from the numerical simulation.}
\label{fig_7_resub}
\end{figure}

\begin{figure}
\includegraphics[width=\columnwidth]{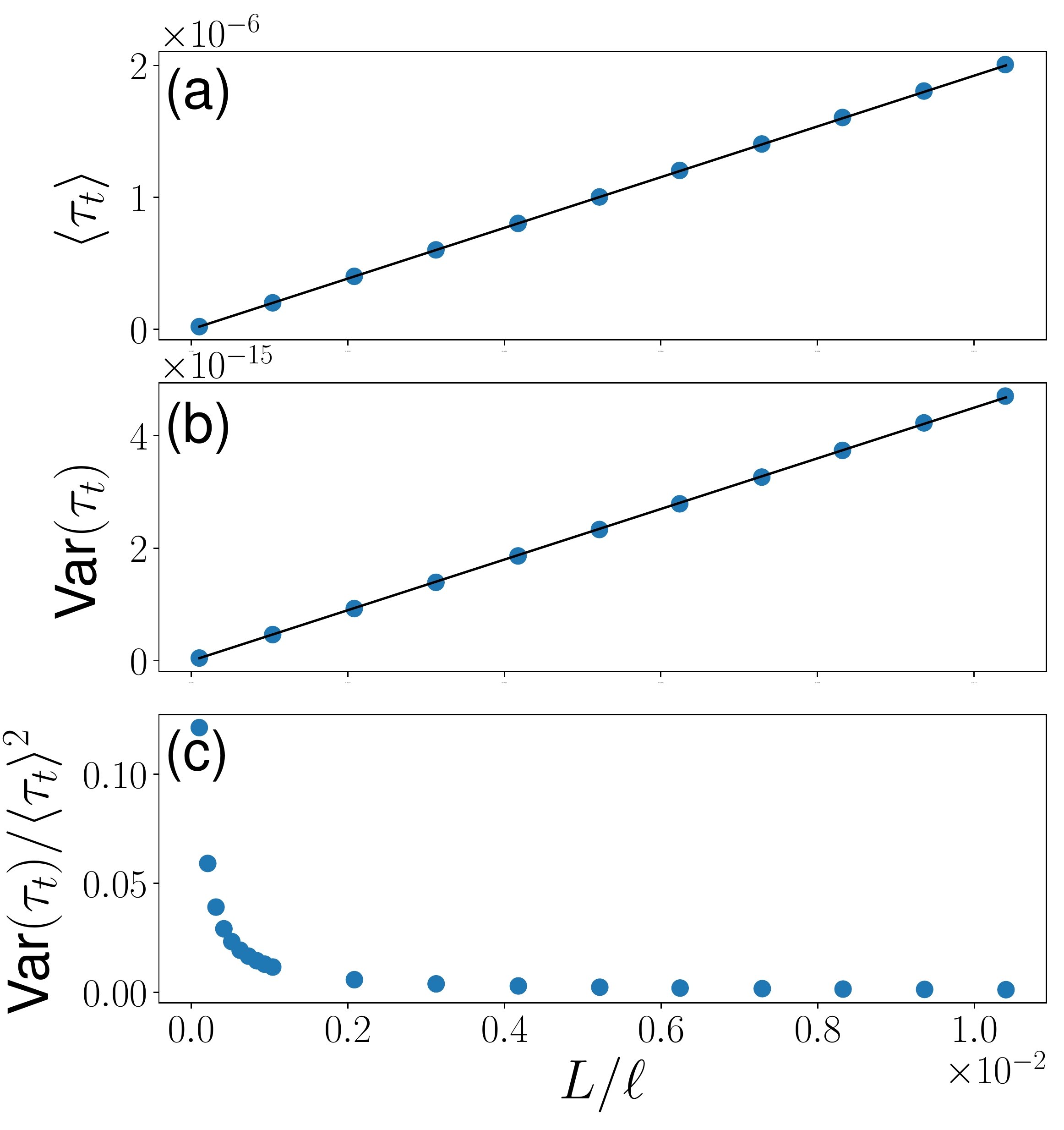}
\caption{(a) Numerical average of the transmission delay-time  for samples with fixed  $\ell\approx 57635$~cm (dots).  The black continuous line is given by  Eq.~(\ref{avertau}). (b) Numerical variance of the reflection delay-time Var$(\tau_t)$  (dots).  The black continuous  line is a fitting of $f(L)=aL$ with $a\approx 7.8\times 10^{-18}$. (c) The ratio $\mathrm{Var}(\tau_t)/\langle \tau_t \rangle^2$ decays to zero as $L$ increases since $\tau_t$ self-averages.}
\label{fig_8_resub}
\end{figure}

In Fig. \ref{fig_7_resub}, we show an example of the distribution of the time delay in transmission, $p(\tau_t)$. 
The average and the variance that characterize the Gaussian distribution  are obtained from the numerical simulations. A good agreement between  the Gaussian distribution (continuous line) and the histogram is seen. 
Additionally, the linear dependence of both the average $\langle \tau_t \rangle$ and variance $\mathrm{Var}(\tau_t)$ with the system length is shown in Figs.~\ref{fig_8_resub}(a) and \ref{fig_8_resub}(b), respectively. The average $\langle \tau_t \rangle$ is given as in Eq.~(\ref{avertau}), since $\langle \tau_t \rangle=\langle \tau_r \rangle$, as previously mentioned. 

In contrast to the time delay in reflection studied in the previous section,  the time delay in transmission is a self-averaging quantity.  
This is shown in Fig. \ref{fig_8_resub}(c), where it is seen that $\mathrm{Var}(\tau_t)/ \langle \tau_t \rangle^2 \to 0$ with $L$, instead of the constant behavior with $L$ shown in Fig. \ref{fig_5_resub}(c).

\section{Summary}
We have presented a statistical analysis of the random fluctuations of the time delay of reflected and transmitted waves in 1D disordered media with high transmission coefficients. 
We have performed numerical simulations of microwaves diffusing in  structures with randomly placed weak scatterers and found that the distribution of the time delay in reflection  
is given by Eq.~(\ref{poftaur}), while the time delay in transmission follows a Gaussian distribution.

We have shown numerically that the random fluctuations of the reflection amplitudes are described by circular complex  Gaussian random variables. This allowed us to apply previous results of~\cite{Genack1999,Tiggelen1999} in which the random field amplitudes were assumed to follow a complex Gaussian distribution. The Gaussian distribution in Eq.~(\ref{Gaussian}) can be seen as consequence of the CLT by considering that the complex field amplitudes $\bf r$ is the  result of a sum of a large number of partial contributions of waves scattered off many randomly-placed scatterers. 

Interestingly enough, the statistics of   
 the time delay in reflection are described as in the diffusive regime, despite that our 1D structures are in the ballistic regime. This does not happen to all other transport quantities. For instance, the distribution of the total transmission in the diffusive and ballistic regimes are completely different. Other unexpected similarities in the ballistic and diffusive regimes  
 such as the same scaling of transmission and intensity channels in both regimes have been found experimentally and numerically \cite{Zhou}.

For the time delay in transmission, we have shown that in a disordered sample, the time delay is given by the sum of partial contributions of random time delays associated with each scattering unit of the samples. Thus, the fluctuations of the time delay in transmission are expected to be described by a Gaussian distribution invoking the CLT. 

 Thus, the statistics of the time delay of both reflected and transmitted waves find their roots in a fundamental result: the central limit theorem. 

From a practical point of view, with the  advances in manufacturing materials, samples with high transmittance have been produced; even in nature, highly transparent materials can be found. However, they are not entirely free of the presence of sources of disorder. Therefore, a statistical analysis  of wave propagation in such materials is of relevance. Our results, however, are strictly valid for 1D systems with high transmission; thus, an extension to higher dimensions is of interest as well as to explore the case of moderated transmission values. In this respect, in Appendix A we analyze the statistical deviations of the reflection amplitudes from a description given by circular Gaussian variables, when the ballistic restriction in our calculations is relaxed.

\begin{acknowledgement}

This work is supported  by MCIU (Spain) under the Project number PGC2018-094684-B-C22. 
L.A.R.-L. acknowledges the financial support by CONACyT through the Grants No. 490639 and No. 775585 (Mexico).
J.A.M.-B. acknowledges support from CONACyT Fronteras Grant No. 425854 (Mexico).

\end{acknowledgement}

\appendix

\section{Deviations of the reflection amplitudes from the Gaussian circular ensemble}

In the main text we have assumed that the reflection amplitudes are complex Gaussian random variables in the ballistic regime. Here we quantify the range of validity of this assumption as a function of the ratio $L/\ell$. 
We perform chi-square goodness of fit tests to determine whether 
the real and imaginary parts of the reflection amplitudes follow a Gaussian distribution and the distribution of $\tau_r$ is described by Eq.~(3), in the main text. 

\begin{figure*}
\includegraphics[width=0.7\textwidth]{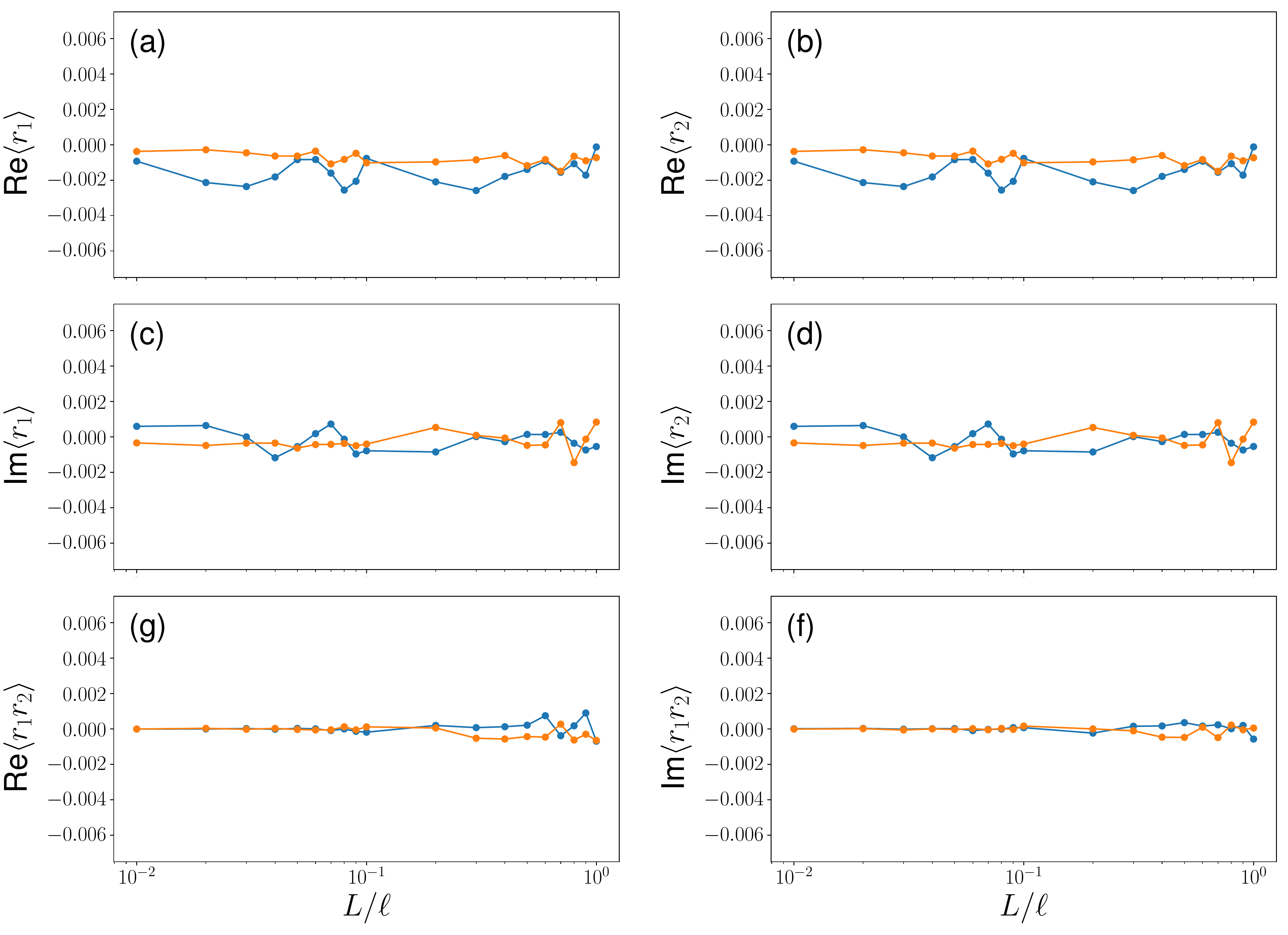}
\caption{Average of the real and imaginary parts of the reflection at two different frequencies ($\omega_1=50.262$~rad ns$^{-1}$ and $\omega_2=50.268$~rad ns$^{-1}$) as a function of the ratio $L/\ell$. Blue and orange dots correspond to $\ell=57634$ cm and 214082 cm, respectively.}
\label{fig_1_appendix}
\end{figure*}

First, we have found that the expectation values that define the circular symmetry, $E[\mathrm{Re}(r_1)]=E[\mathrm{Im}(r_1)]=E[\mathrm{Re}(r_1 r_2)]=E[\mathrm{Im}(r_1 r_2)]=0$, of the reflections are preserved even for values of $L \sim \ell$, as it can be seen in Fig.~\ref{fig_1_appendix}.

\begin{figure*}
\includegraphics[width=0.7\textwidth]{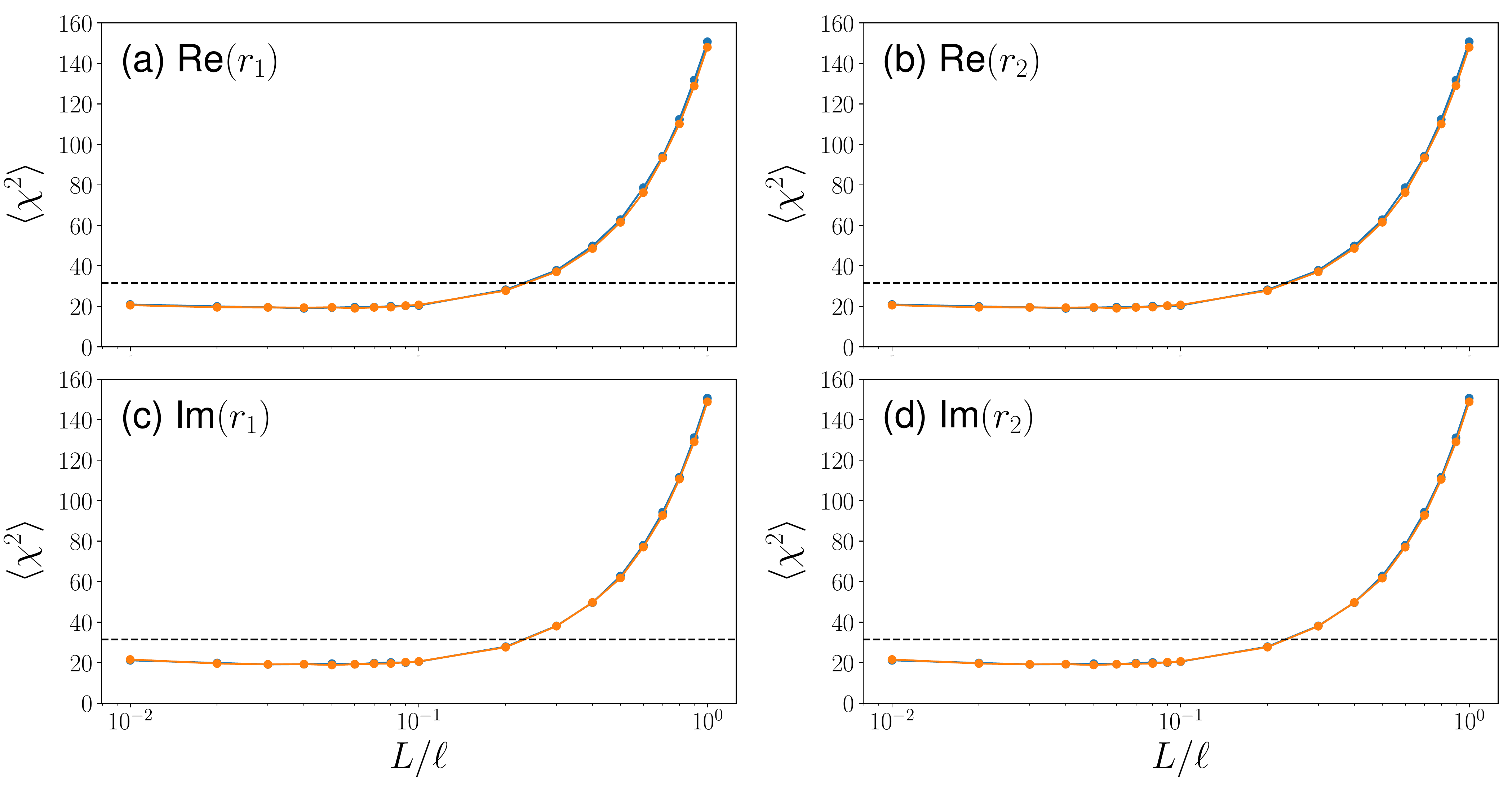}
\caption{Chi-square test values as a function of the ratio $L/\ell$ to test the Gaussian distribution of the real and imaginary parts of the reflection at two frequency values. Blue and orange dots correspond to $\ell = 57634$ cm and 214082 cm, respectively. The dashed line indicate the critical value $\chi^2_c =31.41$ which corresponds to the level of significance $\alpha$ = 0.05 with 20 degrees of freedom.}
\label{fig_2_appendix}
\end{figure*}

However, the chi-square goodness of fit reveals that the reflection amplitudes are no longer described by a Gaussian distribution as $L$ approaches  $\ell$. 
In Fig. \ref{fig_2_appendix}, we show the values of $\chi^2$ of the real and imaginary parts of the reflection at two different frequencies and two values of the mean free path as function of the system length. 
The $\chi^2$ values are obtained from 1000 samples and the data is grouped into 21 histogram classes. We obtained 1000 values of $\chi^2$ for each value of $L/\ell$. We thus plot the average value 
$\langle \chi^2 \rangle$. The horizontal dashed lines in Fig.~\ref{fig_2_appendix} indicate the  $\chi^2$ critical value with 20 degrees of freedom and level of significance  $\alpha$ = 0.05. As it can be seen, the $\chi^2$ values are within the acceptance level up to  $L/\ell \sim 10^{-1}$.

\begin{figure}
\includegraphics[width=0.95\columnwidth]{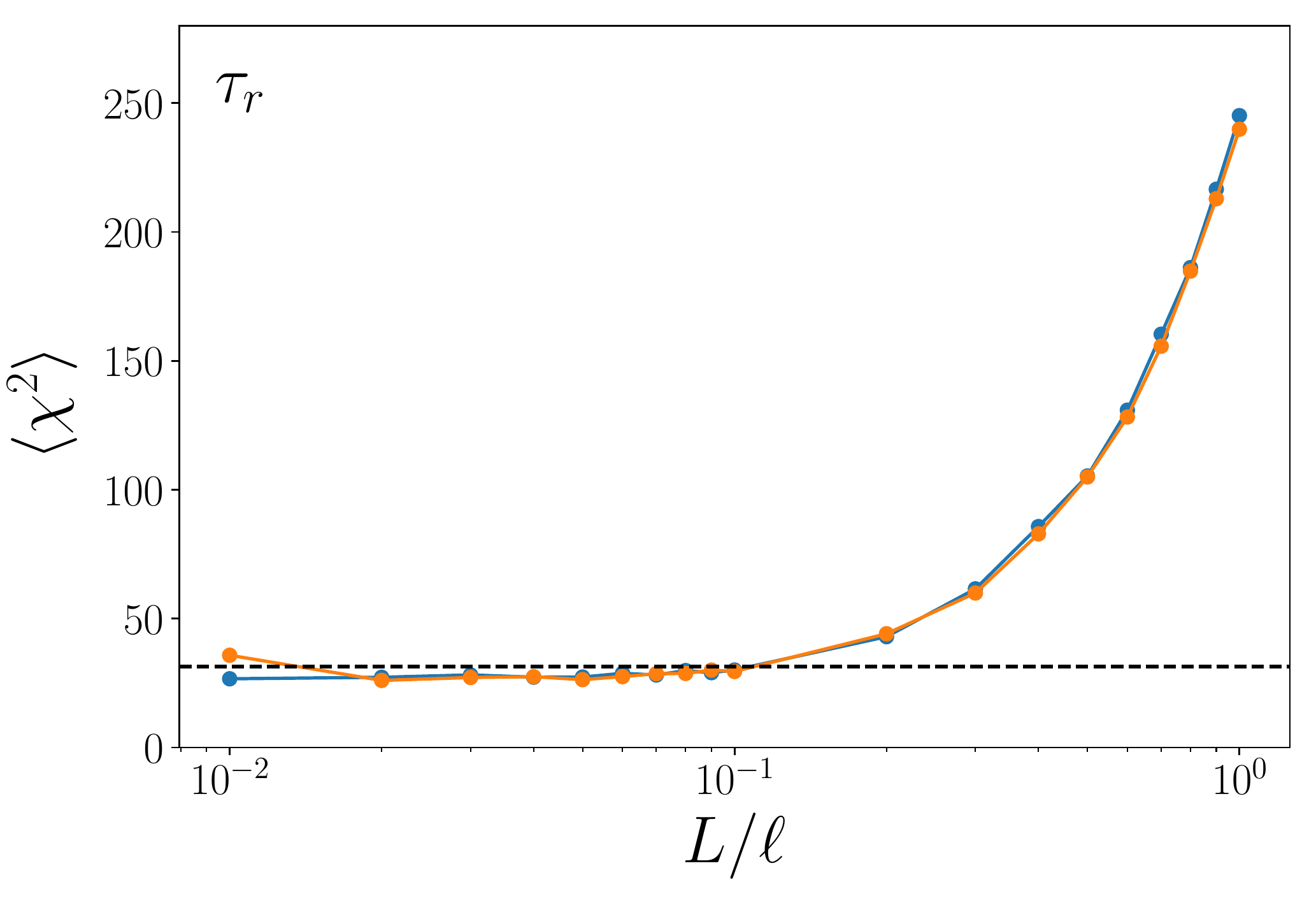}
\caption{Chi-square test values as a function of the ratio $L/\ell$ to test the agreement of the numerical and the expected (Eq. (3)) distribution of $\tau_r$. Blue and orange dots correspond to $\ell=57634$ cm and 214082 cm, respectively. The dashed line indicates the critical value $\chi^2_c =31.41$ which corresponds to the level of significance $\alpha$ = 0.05 with 20 degrees of freedom.}
\label{fig_3_appendix}
\end{figure}

It is thus expected that for systems with $L \sim \ell$, the distribution of the time delay in reflection would not be well described by Eq.~(3) in the main text. Indeed, 
we have performed chi-square tests for the distribution of $\tau_r$.  In Fig.~\ref{fig_3_appendix}, the values of the chi-square goodness of fit indicate that the numerical distributions of $\tau_r$ are well described by Eq. (3) in the main text up to $L/\ell \sim 10^{-1}$.

\end{document}